\documentclass[aps,prl,twocolumn,superscriptaddress,groupedaddress]{revtex4}  
\usepackage{graphicx}  
\usepackage{dcolumn}   
\usepackage{bm}        
\usepackage{amssymb}   
\usepackage{epstopdf}
\usepackage{amsmath}
\usepackage{color}
\usepackage{multirow}
\usepackage{hyperref}

\begin{document}

\title{Experimental test of Mermin inequalities on a five qubit quantum computer}

\author{Daniel Alsina}
\author{Jos\'e Ignacio Latorre}
\affiliation{Dept. F\'isica Qu\`antica i Astrof\'isica, Universitat de Barcelona, Diagonal 645, 08028 Barcelona, Spain.\\} 
\date{\today}

\begin{abstract}
Violation of Mermin inequalities is tested on the five qubit IBM quantum computer. 
For 3, 4 and 5 parties, quantum states that violate the corresponding Mermin
inequalities are constructed using quantum circuits on superconducting qubits. Measurements on different basis
are included as additional final gates in the circuits. The experimental results
obtained using the quantum computer show violation of all Mermin inequalities, with a clear
degradation of the results in the 5 qubit case. 
Though this quantum computer  is not competitive to test Mermin inequalities as compared to other
techniques when applied to few qubits, it does offer the opportunity to
explore
multipartite entanglement for four and five qubits beyond the reach of other alternative technologies.

\end{abstract}

\maketitle

Quantum physics can be discriminated from classical physics using Bell-type inequalities 
\cite{bell}. In particular, the violation of Bell inequalities for two qubits has been extensively verified
since they were first checked in  atomic physics experiments \cite{aspect,all}. Later
on, the improvement of quantum optics techniques as well as other technologies such as NV-centers has
made it possible to eliminate many of the loopholes in the  experimental verification of 2-qubit Bell inequalities \cite{all}.

An extension of Bell inequalities to a larger number of particles corresponds to the set
of Mermin inequalities \cite{mermin}. Such inequalities should be maximally violated by
GHZ-type states \cite{werner}. The experimental verification of multipartite Mermin inequalities faces
the problem of a good control of three or more qubits, including the generation of entangled
states, and the possibility of performing different measurements on each one. Violation
of Mermin inequalities has  been reported for three qubits \cite{mermin-3qubits} and four qubits
\cite{mermin-4qubits}, where all qubits are made out of photons, and for up to 14 qubits with a quantum computer based on ion traps \cite{merminmany}.

In the case of superconducting qubits,
violation of the CHSH inequality was
achieved in \cite{ansmann}, whereas the GHZ construction and the 3-qubit Mermin inequality violation was demonstrated by \cite{dicarlo}.
For a general review of theoretical and experimental progress in Bell inequalities, see \cite{Brunner}.

The construction of the first prototypes of quantum computers allows for the possibility of experimenting
with quantum states containing more than 2 qubits. In particular, IBM has
opened the use of its 5-qubit quantum computer to the community \cite{IBM}. 
 We here shall report results on the use
of this quantum computer to test the violation of Mermin inequalities for 3, 4 and 5 
superconducting qubits.

\section*{Mermin polynomials}

Local realism can be tested using Mermin polynomials. The technique to generate them is explained for example in \cite{Alsina}. The Mermin polynomial for 3-qubits is
\begin{equation}
M_3 = (a_1 a_2 a'_3 + a_1 a'_2 a_3 + a'_1 a_2 a_3) - (a'_1 a'_2 a'_3) \, ,
\label{m3}
\end{equation}
where $a_i$ and $a'_i$ correspond to two different settings for the measurement of each qubit $i$.  
Each  measurement can take the values $\{-1,1\}.$ Classical theories
obey local realism (LR) which translates into a bound for the expectation value of the Mermin polynomial,  $\langle M_3 \rangle ^{LR} \leq 2 $.
Instead, for quantum mechanics (QM) the observables $a_i$ and $a'_i$ are  built out of linear combinations of Pauli matrices. Each
measurement is expressed as a Kronecker product of the three local measurements and the
expectation value for $\langle M_3 \rangle $ is  the maximum eigenvalue of the resulting 8x8 matrix. 
In this case, the maximum possible eigenvalue, and therefore the quantum bound, is $\langle M_3 \rangle ^{QM} \leq 4$.
We shall shortly construct circuits to check the violation of the classical bound on this inequality.

The  Mermin polynomial that will be experimentally checked for 4-qubits is
\begin{eqnarray}
\label{m4}
M_4 =&& -(a_1 a_2 a_3 a_4) \\ \nonumber 
&&+ (a_1 a_2 a_3 a'_4 + a_1 a_2 a'_3 a_4 + a_1 a'_2 a_3 a_4 + a'_1 a_2 a_3 a_4) \\ \nonumber
&&+ (a_1 a_2 a'_3 a'_4 + a_1 a'_2 a_3 a'_4 + a_1 a'_2 a'_3 a_4  \\ \nonumber 
&&+ a'_1 a_2 a_3 a'_4 + a'_1 a_2 a'_3 a_4 + a'_1 a'_2 a_3 a_4) \\ \nonumber 
&&- (a_1 a'_2 a'_3 a'_4 + a'_1 a_2 a'_3 a'_4 + a'_1 a'_2 a_3 a'_4 + a'_1 a'_2 a'_3 a_4) \\ \nonumber 
&& -(a'_1 a'_2 a'_3 a'_4)\nonumber, ,
\end{eqnarray}
with a classical bound of $\langle M_4 \rangle ^{LR}\leq 4$ and a quantum bound of $ \langle M_4 \rangle ^{QM} \leq 8\sqrt{2}$\hspace{1mm}.

In the 5-qubit case, the Mermin polynomial reads
\begin{eqnarray}
\label{m5}
M_5 =&& -(a_1 a_2 a_3 a_4 a_5) \\ \nonumber 
&&+ (a_1 a_2 a_3 a'_4 a'_5 + a_1 a_2 a'_3 a_4 a'_5 + a_1 a'_2 a_3 a_4 a'_5 \\ \nonumber 
&&+ a'_1 a_2 a_3 a_4 a'_5 + a_1 a_2 a'_3 a'_4 a_5 + a_1 a'_2 a_3 a'_4 a_5 \\ \nonumber
&&+ a'_1 a_2 a_3 a'_4 a_5 + a_1 a'_2 a'_3 a_4 a_5 + a'_1 a_2 a'_3 a_4 a_5 \\ \nonumber 
&&+ a'_1 a'_2 a_3 a_4 a_5) \\ \nonumber
&&-(a_1 a'_2 a'_3 a'_4 a'_5 + a'_1 a_2 a'_3 a'_4 a'_5 + a'_1 a'_2 a_3 a'_4 a'_5
\\ \nonumber
&& + a'_1 a'_2 a'_3 a_4 a'_5 + a'_1 a'_2 a'_3 a'_4 a_5)\, ,
\end{eqnarray}
with a classical bound of $\langle M_5 \rangle ^{LR}\leq 4$ and a quantum bound of $\langle M_5 \rangle ^{QM} \leq 16$.

\section*{Circuit implementation}

There are a number of technical issues associated to the specific implementation of the IBM 5-qubit quantum
computer. This quantum computer
is based on superconducting flux qubits that live on a fridge with a temperature of about 15 mK, where only one of the qubits can be used to act as the target qubit of any CNOT gate.
In the test of Mermin inequalitites, only  GHZ-like states have to be created. This requires the use of a Hadamard gate on a control
qubit followed by CNOTs targeted to the rest. In order to implement this kind of action we shall need to operate
CNOT gates targeted to other qubits. This can be done using the relation
$CNOT_{1\to 2} =(H_1 \otimes H_2) CNOT_{2\to 1} (H_1 \otimes H2)$, where $H_1$ and $H_2$ are Hadamard
gates on qubits 1 and 2, whereas $CNOT_{1\to 2}$ is the control-NOT gate which is controlled by qubit 1.\\

In our choice of settings, the needed GHZ-like states have relative phases, as in the case of 3-qubits, where $|\phi\rangle=1/\sqrt{2}(|000\rangle+i|111\rangle)$. These phases are implemented using S and T gates, which are one-qubit gates that mutiply the $|1\rangle$ term with $\pi/2$ and $\pi/4$  phases, respectively.
Measurements can only be done on the $\sigma_z$ basis, but they can be simulated in another basis with the help of additional gates, namely an H gate for $\sigma_x$ and an $S^\dagger$ gate followed by an H gate for $\sigma_y$. \\

Another relevant issue to be considered is that not all of the qubits are equally robust in the present
quantum computer, some have relaxation and decoherence times larger than others, although all of them are of the order of $T = \mathcal{O} (100 \mu s)$. We shall adapt our circuits to minimize the  number of gates on
the qubits that behave more poorly. For example gates that implement phases that can be put freely in any qubit are allocated to the most robust ones. \\

Figures 1 and 2 represent the three circuits for the 3, 4 and 5 qubit Mermin inequalities. In principle we need to perform as many experiments as the number of terms in the Mermin inequalities \eqref{m3}, \eqref{m4} and \eqref{m5}. However due to our limited access to the computer and the symmetry of particle exchange of the states and the inequalities, only one experiment for a term representative of each number of primes $(a'_i)$ is run. In our choice of settings, the number of primes amounts to the number of $\sigma_y$ measurements, whereas the non primes ($a_i$) correspond to $\sigma_x$ measurements.
 We thus have 2 experiments for 3-qubits, 5 experiments for 4-qubits and 3 experiments for 5-qubits. Each experiment is run 8192 times, the maximum available, except for the 3-qubit experiments, which have been run only 1024 times. When computing the expected value of the whole polynomial, each experiment is given the corresponding weight. In the errors discussion we compare results obtained when using the symmetry with results obtained without using it, computing all the terms, for the three-qubit case.

\begin{figure}[h!]
\centering
\includegraphics[scale=0.35]{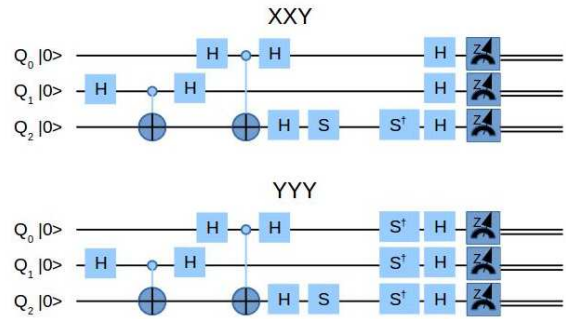}
\label{circuit3}
\caption{The two circuits used for the three-qubit Mermin inequality. The first circuit corresponds to $\sigma_x\sigma_x\sigma_y$ experiment, and the second circuit to $\sigma_y\sigma_y\sigma_y $ experiment. The $S^\dagger$ gates make the difference between a $\sigma_x$ and a $\sigma_y$ measurement.}
\end{figure}

\begin{figure}[h!]
\centering
\includegraphics[scale=0.35]{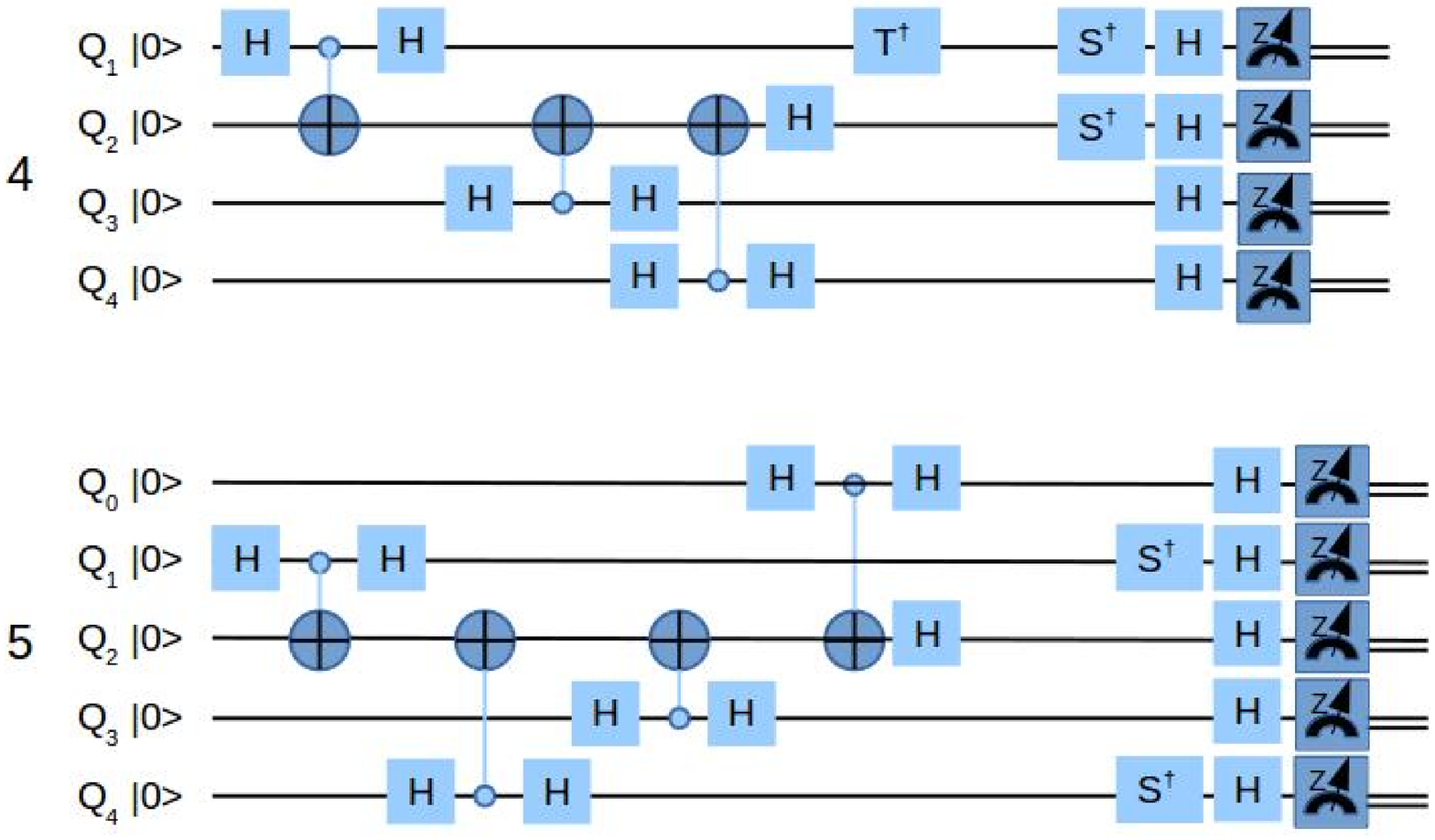}
\label{circuit45}
\caption{Two of the circuits used for the four-qubit and five-qubit Mermin inequalities. The first circuit corresponds to $\sigma_y\sigma_y\sigma_x\sigma_x$ experiment, whereas the second corresponds to $\sigma_x\sigma_y\sigma_x\sigma_x\sigma_y $ experiment. The $S^\dagger$ gates make the difference between a $\sigma_x$ and a $\sigma_y$ measurement. In order to change from $\sigma_x$ to $\sigma_y$, one has to add an $S^\dagger$ gate, or remove it to do the opposite. With this technique one can obtain all circuits needed to test the inequalities. }
\end{figure}

\section*{Results}

We shall now give a more detailed discussion of the results for the 3-qubits case and an abridged one for the 4 and 5-qubit cases, as much of it is basically the same.

In order to check the violation of the inequality, one has to choose the settings and the corresponding state that maximally violate it. One possibility is to choose settings $a_i=\sigma_x$ and $a'_i = \sigma_y$ for all the qubits. The state that maximizes the quantum violation in this case is $|\phi\rangle=1/\sqrt{2}(|000\rangle+i|111\rangle)$.

The 3-qubit Mermin inequality has 4 terms as shown in Eq. \eqref{m3}. In principle,  four different circuits are needed, one for each term. The state will be the same for all of them, but the settings change. However, one can use the symmetry of the state and the inequality to reduce the number of measurements needed if there is limited access to the experimental setting as is our case. All the terms that have the same number of primes $(a'_i)$ are represented by the same circuit by symmetry.   We then considered only two different experiments, with 1024 runs each, the $\sigma_x \sigma_x \sigma_y$ experiment and the $\sigma_y \sigma_y \sigma_y$ experiment. The results are shown in table \ref{results3} .

\begin{table}[h!]
\centering
\resizebox{\linewidth}{!}{%
\begin{tabular}{c | c | c | c | c | c | c | c | c}
\hline
Result XXY & \textbf{000} & \textit{001} & \textit{010} & \textbf{011} & \textit{100} & \textbf{101} & \textbf{110} & \textit{111} \\
Probability & 0.229 & 0.042 & 0.024 & 0.194 & 0.043 & 0.203 & 0.231 & 0.033 \\
\hline
Result YYY & \textbf{000} & \textit{001} & \textit{010} & \textbf{011} & \textit{100} & \textbf{101} & \textbf{110} & \textit{111}   \\
Probability & 0.050 & 0.188 & 0.188 & 0.028 & 0.258 & 0.026 & 0.041 & 0.221 \\
\end{tabular}%
}
\caption{Table of detailed results for the two 3-qubit experiments. In bold are results of even parity, in italic results of odd parity. Counts for each result are expressed in probabilities computed out of 1024 runs. Computation of the expected value of XXY gives $\langle XXY \rangle = 0.715$ and for YYY gives $\langle YYY \rangle = -0.710.$ The combination 3  $\langle XXY \rangle - \langle YYY \rangle$ gives $\langle M_3 \rangle_{exp} = 2.85\pm 0.02$. }
\label{results3}
\end{table}

Eight probabilities for each term are obtained. In order to translate these probabilities to the expected values that appear in the inequality, one has to arrange the results in two groups according to the parity of the number of 1 (which represent the value -1.) The expected value of the term is obtained by summing all the probabilities of the results of even parity and substracting the results of odd parity. The correctly weighted sum of the expected values of each term gives the final result $\langle M_3 \rangle_{exp} = 2.85\pm 0.02$.

In the case of 4-qubits, the use of settings $a_i=\sigma_x$ and $a'_i = \sigma_y$ implies that the state that maximizes the quantum violation is $|\phi\rangle=1/\sqrt{2}(e^{ i \pi/4}|0000\rangle+|1111\rangle)$. With these settings and this state, 5 experiments are performed, one for each term with different number of primes \eqref{m4}, with 8192 runs for each experiment. A result of $\langle M_4 \rangle_{exp} = 4.81\pm 0.06$ has been obtained.

In the case of 5-qubits, the use of settings $a_i=\sigma_x$ and $a'_i = \sigma_y$ implies that the state that maximizes the quantum violation is $|\phi\rangle=1/\sqrt{2}(|00000\rangle+|11111\rangle)$. With these settings and this state, 3 experiments are performed, one for each term with different number of primes \eqref{m5}, with 8192 runs for each experiment. A result of $\langle M_5 \rangle_{exp} = 4.05\pm 0.06$ has been obtained. This is clearly a poor violation, which is still compatible
with local realism. Improvement of the quantum computer is needed to obtain more accurate results.

\begin{table}[h!]
\centering
\begin{tabular}{c | c | c | c | c | c |}

     & LR & QM & EXP  \\
\hline
3 qubits & 2 & 4 & \textbf{2.85$\pm$ 0.02}   \\
4 qubits & 4 & 8 ${\sqrt 2}$ & \textbf{4.81$\pm$ 0.06}   \\
5 qubits & 4 & 16 & \textbf{4.05$\pm$ 0.06}   \\

\end{tabular}
\caption{Table of results. LR corresponds to the local realism bound for each Mermin  inequality, QM to the quantum bound and EXP is the experimental result. }
\label{tabresults}
\end{table}

The results obtained from the IBM quantum computer are subject to
different kind of errors.

The stability of the quantum computer is still poor
and the same experiments run at different times provide results that differ more than the expected
behaviour of statistical fluctuations. As an example, one month after the original runs, the 3-qubit experiment has been run again to compare results. This time, a result of $\langle M_3 \rangle_{exp} = 2.57\pm 0.02$ has been obtained, clearly showing the previous point. An additional run has been done computing separately the four terms of \eqref{m3}, without assuming any symmetry, and a similar result is obtained, $\langle M_3 \rangle_{exp} = 2.57\pm 0.02$, showing that it is safe to assume the symmetry of party exchange.

We may get an estimation of the statistical error as a dispersion
around the mean. We may, as well, treat the results as a multinomial
distribution, using the expression $\delta p = \sqrt{p(1-p)/N}$, which for N=8192 gives $\delta p = \mathcal{O} (10^{-2})$. The different Mermin inequalities for 3, 4 and 5 qubits
require a different number of experiments to be done, which are
considered as independent. We may then add in quadrature its errors,
which is the figure we associate in the explicit results. In this sense, the 5-qubit
result obtained in the present quantum computer does not have sufficient statistical
significance to discard local realism.

Furthermore, some of the issues related to the elimination of loopholes can
not be addressed. Experiments suffer from errors
related to stability,
loss of coherence and lack of full fidelity of the quantum gates. This is
clearly seen as the
violation of Mermin inequalities will deteriorate progressively as
the numbers of qubits,
and gates used in the experiment, increase. We may think of the
experimental verification
of Mermin inequalitites as a test of the overall fidelity of the whole 
Mermin circuits.

\section*{Conclusions}

Experimental verification of Mermin inequalities for 3, 4 and 5 qubits
has been tested
on a 5 superconducting qubit IBM quantum computer. Results do show violation
of local realism
in all cases, with a clear degradation in quality as the number of
qubits (and needed gates)
increases. Nonetheless, this produces the first experimental violation
of 4 and 5-qubit Mermin inequalities with superconducting qubits, though the statistical significance of the second one is still poor.
still poor. It should be noted however that in the case of the 4-qubit inequality, the result shows generic non-locality but does not provide evidence for genuine four-particle non-locality, because this would only be implied by $M_4 > 8$. \cite{Collins}.
It can be argued that the measurements of Mermin polynomials for many
qubits can be used as
a figure of merit to assess the fidelity of a quantum computer.

\section{Acknowledgements}

D.A. acknowledges financial help from the APIF Scholarship of University of Barcelona. 
JIL acknowledges financial support
by FIS2013-41757-P.
We acknowledge use of the IBM Quantum Experience for this work. The views expressed are those of the
authors and do not reflect the official policy or position of IBM or the IBM Quantum Experience team.

\end{document}